\documentclass[prb, showpacs, floats, superscriptaddress, twocolumn]{revtex4-1}

\usepackage[utf8x]{inputenc}
\usepackage{amsfonts}
\usepackage[T1]{fontenc}
\usepackage{bbold}
\usepackage{amsmath}    
\usepackage{graphicx}   
\usepackage{verbatim}   
\usepackage{color}      
\usepackage{url}
\usepackage{natbib}
\usepackage[colorlinks,bookmarks=false,citecolor=blue,linkcolor=blue,urlcolor=blue, linktoc=page]{hyperref}

\newcommand{\sk}{\sqrt{K}}
\newcommand{\pp}{\text{:}} 
\newcommand{\zo}{\overline{z}} 
\newcommand{\gt}{\gamma_{2p}} 
\newcommand{\gtp}{\gamma_{2p}^{in}} 
 
\newcommand{\De}{D_{\eta}} 
\newcommand{\DeT}{\tilde{D}_{\eta}} 
\newcommand{\uT}{\tilde{u}} 
\newcommand{\udT}{\tilde{u}_2}

\newcommand{\mt}{\tilde{m}}

\begin{document}
\title{Random Rashba spin-orbit coupling at the quantum spin Hall edge}

\author{Florian Geissler}
  \affiliation{Institute for Theoretical Physics and Astrophysics,
 University of W\"urzburg, 97074 W\"urzburg, Germany}
\author{Fran\c{c}ois Crépin}
  \affiliation{Institute for Theoretical Physics and Astrophysics,
 University of W\"urzburg, 97074 W\"urzburg, Germany}
\author{Bj\"orn Trauzettel}
  \affiliation{Institute for Theoretical Physics and Astrophysics,
 University of W\"urzburg, 97074 W\"urzburg, Germany}

\date{\today}

\begin{abstract}

We study a one-dimensional helical system with random Rashba spin-orbit coupling. Using renormalization group methods, we derive a consistent set
of flow equations governing the important control parameters of the backscattering process. Thereby, we prove the existence of disorder-induced two-particle backscattering that can even be non-local in space. This analysis allows us to derive the scaling form of the conductance at low temperatures.
We find that two-particle backscattering due to random spin-orbit coupling differs from the one off a single Rashba impurity by both the scaling of the conductance with the temperature and 
the relevance of the backscattering operators.

\end{abstract}


\maketitle

\section{Introduction}

A quantum spin Hall state is a topologically non-trivial state of matter exhibiting an energy gap in the bulk. This topological phase is characterized by gapless edge channels that give rise to peculiar transport properties.~\cite{KaneMele, KaneMele2,Ber06} Importantly, edge electrons form a one-dimensional (1D) helical liquid, where the (pseudo) spin degree of freedom is strongly coupled to the direction of motion. The conducting edge channels (with a linear energy dispersion) always come in counter-propagating pairs that are time-reversed partners. Such 1D helical liquids have been experimentally realized at the edges of two-dimensional quantum spin Hall insulators such as HgTe/CdTe \cite{Mole} or InAs/GaSb quantum wells.~\cite{Kne11}

Since elastic backscattering off non-magnetic impurities is prohibited within the edge states by time reversal symmetry (TRS), such helical systems give prospect of robust ballistic electronic transport. To quantify this robustness, it is important to better understand possible sources of backscattering and the influence of disorder on transport properties of helical liquids. Soon after the first prediction of helical edge states, it was realized that their transport properties can be affected by \textit{inelastic} single-particle or multi-particle backscattering.~\cite{Xu06,Wu06} These processes require external scattering potentials that enable to spin-flip the backscattered electron into its counter propagating channel. This can, for instance, be done by a local variation of Rashba spin-orbit coupling (SOC) in the presence of electron-electron interactions.~\cite{Stroem}

The influence of a single scatterer on the transport properties of an interacting helical liquid has been studied by various groups under different assumptions. \cite{Jan,Sch12,Crepin1,Oreg,Vay13} Additionally, backscattering off a Kondo impurity \cite{Mac09, Ta11} or dynamically ordered nuclear spins \cite{Mae13} has also been addressed. All these works have in common that they predict a particular temperature dependence of edge channel transport (typically a power-law behavior) which has thus far not been seen in experiments. However, it is fair to say that experiments have not yet carried out a careful analysis of the temperature dependence of transport. Thus, more experiments on cleaner systems at a wide temperature range are needed to clarify the role of inelastic scattering in helical liquids.

\begin{figure}[htb]
\includegraphics[width=0.45\textwidth]{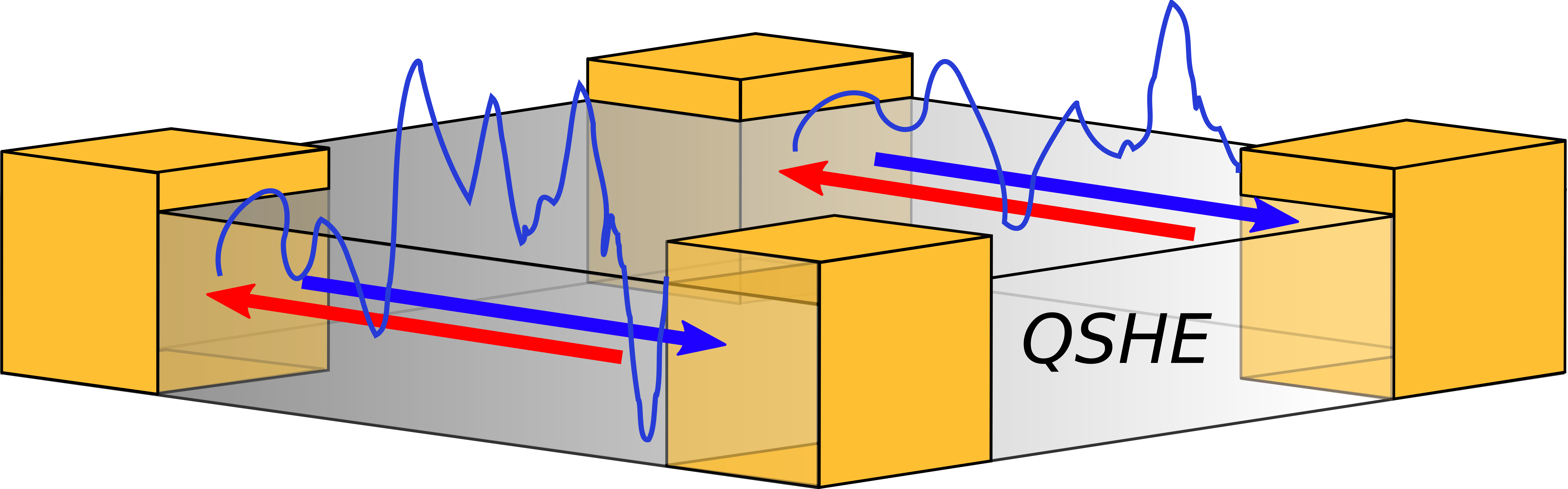}
\caption{(Color online) Experimental setup: A 2D topological insulator in a quantum spin Hall state (gray layer) is connected to four leads. At the edges, this gives rise to conducting 1D helical edge modes.
Disorder is now accomplished by a random potential $\alpha$ (blue wavy lines), originating e.g. from an external source, surrounding layers or interspersed impurities.}
\label{fig:PlotSetup}
\end{figure}

Moreover, the influence of many impurities on edge channel transport in quantum spin Hall systems is much less understood. Previous theoretical work \cite{Xu06,Wu06,Stroem} mainly aimed to map this problem onto the known problem of Anderson localization in an ordinary 1D Luttinger liquid.~\cite{GiaPa}
In this article, we show that such a mapping is not simply achievable in the presence of random Rashba SOC. Instead, the physics of the disordered helical edge states, as schematically shown in Fig.~(\ref{fig:PlotSetup}), is much richer.
This statement is in accordance with a recent analysis of the model introduced in Ref. \onlinecite{Sch12} in the presence of uncorrelated disorder \cite{Kain14}.
We use bosonization in combination with a renormalization group (RG) analysis to derive a set of flow equations for the parameters of interest. Thereby, we obtain a correction to the dc conductance due to two-particle backscattering (TPB) which again exhibits a distinct temperature dependence.
The same physics scenario was previously analyzed by Str{\"o}m {\it et al.} in Ref. \onlinecite{Stroem}. However, these authors used a path integral approach to solve the problem which seems to make the wrong prediction that finite backscattering remains in the non-interacting limit. Indeed, we show below that it is highly non-trivial to obtain the correct non-interacting limit in our formalism which is an operator-based RG analysis. Interestingly, we derive that the correction to the conductance $\delta G$ due to random Rashba SOC should scale for large system sizes as
\begin{align}\label{eq:Tscaling}
&\delta G \sim \begin{cases}
		 -T^{4K-1}  &\mbox{if } 1/4<K<1, \\
                 -T^{8K-2}  &\mbox{if } K<1/4,
            \end{cases}
\end{align}
where $T$ is the temperature and $K$ the interaction parameter that characterizes the interaction strength of the Luttinger liquid. Within our approach, we find that the Rashba disorder is actually a relevant perturbation as soon as $K<1/2$ (see Sec. II), which corresponds to rather strong, or even long-ranged Coulomb interactions. It is not obvious that localization, induced by two-particle backscattering, should occur at low energies, in contrast to usual disordered Luttinger liquids. There indeed, the strong-coupling region contains the free-fermion limit, which exhibits Anderson localization~\cite{GiaPa}. Nevertheless, the reader should be aware that the scaling of Eq.~\eqref{eq:Tscaling}, for $K<1/4$, is probably cut off at temperatures below the typical energy controlling the strong-coupling fixed point.

The paper is organized as follows. In Sec.~II, we describe the model including some details about the disorder average. Subsequently, in Sec.~III, we outline the RG calculation and present the relevant flow equations of the system. Particularly, we discuss in detail the possibility of a spatially separated TPB (see Sec.~III.A) and its local approximation (see Sec.~III.B). In Sec.~IV, the temperature dependence of the correction to the conductance is analyzed before we conclude in Sec. V. Some technical details of the operator product expansion (OPE) and the RG calculation are moved to the Appendices.

\section{Model}

Our model describes interacting electrons in a 1D helical liquid in the presence of Rashba SOC. Since the spin direction in a helical liquid is locked to the
direction of motion, electrons can effectively be considered as spinless. The Hamiltonian consists of three terms, $H=H_0+H_I+H_R$, with \cite{Crepin1, Stroem}
\begin{align}
& H_0= \int dx\ \sum_{r=\pm}\Psi^{\dagger}_r(x)\left(-i r v_F \partial_x-E_F \right)\Psi_r(x),\\
& H_I= g_2 \int dx\ \Psi^{\dagger}_+(x) \Psi^{\dagger}_-(x) \Psi_-(x)\Psi_+(x),\\
& H_R=\int dx\ \alpha(x)\left[ \left(\partial_x \Psi^{\dagger}_+\right) \Psi_- - \Psi^{\dagger}_+ \left(\partial_x \Psi^{\phantom{\dagger}}_- \right) \right]+\textrm{H.c..}
\end{align}
$\Psi^{\dagger}_{\pm}(x)$ and $\Psi_{\pm}(x)$ are fermionic creation and annihilation operators for a right $(+)$ or left $(-)$ moving particle, $E_F$ is the Fermi energy, and $v_F$ the Fermi velocity. We set $\hbar=1$ in this article unless explicitly stated.
$H_0$ describes the free Hamiltonian with a strictly linear dispersion relation. $H_I$ embodies electron-electron interactions of a $g_2$ type between electrons propagating in opposite directions. We do not explicitly take into account interactions of the (chiral) type $g_4$, between electrons moving in the same direction since they only renormalize the Fermi velocity~\cite{Gia, Crepin1}. So-called $g_1$ interactions, that backscatter electrons, are generally forbidden in a helical liquid, as the Coulomb potential does not flip spin. $H_R$ describes the Rashba SOC that couples right and left movers~\cite{Stroem} and we take $\alpha(x)$ to be a random function, in order to model disorder. In the following, we will treat interactions exactly and bosonize the fermionic Hamiltonian. We make use of the bosonization identity~\cite{Haldane, Delft}
\begin{equation*}
 \Psi_{\pm}(x)=\kappa_{\pm} \frac{1}{\sqrt{2\pi a}}e^{\pm i k_F x}e^{-i\left(\pm \phi(x) -\theta(x) \right)}\;,
\end{equation*}
where $\kappa_{\pm}$ is the Klein-factor for a right/left moving particle, $a$ a short-distance cutoff, $k_F$ the Fermi momentum, and $\phi$ and $\theta$ two bosonic fields obeying the commutation relation $[\phi(y), \partial_{x}\theta(x)]=i \pi \delta(x-y)$. The Hamiltonian, in its bosonized form, is now given by $H = \mathcal{H}_0 + \mathcal{H}_R$, with  
\begin{align}
\label{eq:H0Def}
 \mathcal{H}_0 & =\frac{v}{2\pi} \int_0^L dx \left[ K \pp(\partial_x \theta)^2\pp+ \frac{1}{K}\pp(\partial_x \phi)^2\pp \right], \\
 \mathcal{H}_R &=i \kappa_+ \kappa_- \int_0^L dx \, \frac{\alpha(x)}{\pi a } \left(\frac{2\pi a}{L}\right)^K \times \notag \\
\label{eq:HRDef}
& \times \left( \pp \partial_x \theta(x)
  e^{-i2 \phi(x)}
 e^{i2k_F x} \pp + \textrm{H.c.}\right)\;
\end{align}
with operators between columns, $\pp (\ldots) \pp$, being normal-ordered. The bosonized Hamiltonian $\mathcal{H}_0$ is the one of a free boson, although it does include all effects of Coulomb interactions, in the values of the plasmon velocity $v$ and the interaction parameter $K$. For repulsive (resp. attractive) interactions, one has $K<1$ (resp. $K>1$), while for free fermions $K=1$ and $v=v_F$.
%
If both $\alpha(x)$ and the fields $\phi(x)$, $\theta(x)$ typically vary only on length scales much bigger than the Fermi wavelength, the integrand of Eq.~\eqref{eq:HRDef} will average out upon integration. This puts constraints on forms of $\alpha(x)$ that lead to non-trivial results. In a helical Luttinger liquid, $\phi(x)$ and $\theta(x)$ are not necessarily slowly varying functions compared to the oscillating factors $e^{\pm i 2 k_F x}$, since the chemical potential may very well be close to the Dirac point $k_F = 0$.  
Away from half filling, one can compensate the factors of $e^{\pm i2 k_F x}$ with the Rashba potential $\alpha(x)$, assuming that the combinations
$\eta(x)=\alpha(x) e^{i2k_F x}$ and $\eta^{*}(x)=\alpha(x) e^{-i2k_F x}$ are now slowly varying. This is the situation we will address in the following.

Treating disorder is a notoriously difficult task, and only a handful of analytical methods are available. The one we chose here is based on the replica trick and has proved to be efficient in the study of 1D interacting electron gases in disordered potentials.  The time-ordered correlation function $\mathcal{A}(\tau_1,\tau_2)=\langle \mathcal{T} O(\tau_1) O(\tau_2) \rangle$ in imaginary time, for an arbitrary observable $O$, for instance, the current density, is given by
\begin{equation}
\mathcal{A}(\tau_1,\tau_2) = \frac{1}{\mathcal{Z}}\textrm{Tr} \left[ e^{-\beta \mathcal{H}_0} \left(\mathcal{T} \; \hat{U}(\beta,0) \hat{O}(\tau_1) \hat{O}(\tau_2) \right) \right] \label{Eq:corr1}
\end{equation}
with $\mathcal{Z}=\mathrm{Tr} [e^{-\beta \mathcal{H}_0}\hat{U}(\beta,0) ]$ the partition function and $\hat{U}(\beta,0)=\mathcal{T} \exp\left[-\int_0^{\beta} d\tau_1 \hat{\mathcal{H}}_R(\tau_1)\right]$ the evolution operator. The average over disorder realizations reads
\begin{align}\label{eq:Averages}
 \overline{\mathcal{A}(\tau_1,\tau_2)}= \int \mathcal{D}\eta\mathcal{D}\eta^*\ p(\eta, \eta^*)
 \mathcal{A}(\tau_1,\tau_2)\;,
\end{align}
where $p(\eta, \eta^*)$ is the probability distribution of the random potential. The partition function in the denominator of Eq.~\eqref{Eq:corr1} makes the average in Eq.~\eqref{eq:Averages} intractable. The replica trick builds on the observation that $\mathcal{Z}^{-1} = \mathcal{Z}^{N-1}$ in the limit $N \to 0$. We then express the denominator using $N-1$ identical (replicated) copies of the system and arrive at
\begin{widetext}
\begin{equation}
\mathcal{A}(\tau_1,\tau_2) =  \lim_{N \to 0} \frac{1}{N}\sum_{a=1}^N \mathcal{A}^{(a)}(\tau_1,\tau_2) = \lim_{N \to 0} \frac{1}{N} \sum_{a=1}^N \textrm{Tr} \left[ e^{-\beta \mathcal{H}_{0,\textrm{rep}}} \left(\mathcal{T} \; \hat{U}_{\textrm{rep}}(\beta,0) \hat{O}^{(a)}(\tau_1) \hat{O}^{(a)}(\tau_2) \right) \right]\;, \label{Eq:corr2}
\end{equation}
\end{widetext}
where $\mathcal{H}_{0,\textrm{rep}} = \sum_{a=1}^N \mathcal{H}_0^{(a)} $ and $\hat{U}_{\textrm{rep}}(\beta,0) = \mathcal{T} \exp\left[ - \sum_{a=1}^N \int_0^{\beta} d\tau_1 \hat{\mathcal{H}}^{(a)}_R(\tau_1)\right]$. For simplicity, we consider a Gaussian probability distribution of the form $p(\eta, \eta^*)=e^{-\De^{-1} \int dx\; \eta^*(x)\eta(x)}$. The disorder strength $\De$ is then the weight of the Gaussian statistics. We assume the random potential to be short-ranged, as $\overline{\eta^*(x_1) \eta(x_2)}=\De \delta(x_1-x_2)$, and with zero mean value, that is, $\overline{\eta(x)} = \overline{\eta^*(x)}= 0$.  Averaging over disorder the quantity $ \mathcal{A}^{(a)}(\tau_1,\tau_2)$ in Eq.~\eqref{Eq:corr2} gives rise to the effective evolution operator $\overline{\hat{U}_{\textrm{rep}}}(\beta,0) \equiv \mathcal{T} \exp \bigg[  \int_0^\beta d\tau_1 \int_0^\beta d\tau_2\ \hat{\mathcal{H}}_{\textrm{dis}}(\tau_1, \tau_2) \bigg]$ with
\begin{widetext}
\begin{align}\label{eq:DefSRrepNew}
\hat{\mathcal{H}}_{\textrm{dis}}(\tau_1, \tau_2)  =
 \frac{\De}{2} \frac{1}{\pi^2 a^2 } \left(\frac{2\pi a}{L}\right)^{2K} \sum_{a,b=1}^N \int_0^L dx  \;\; \pp \partial_x \theta_a(x, \tau_1)  e^{+i2 \phi_a(x, \tau_1)}  \pp\ \times\  \pp \partial_{x} \theta_b(x, \tau_2)  e^{-i2\phi_b(x, \tau_2)}  \pp + \textrm{H.c.}
\end{align}
\end{widetext}
Note that the Klein factors are set equal to one for simplicity throughout this article, since they always come in pairs and do not affect the final results. Our philosophy for the rest of the paper is to first perform an RG analysis of the Rashba potential and uncover a preliminary phase diagram from the flow equations, and to then deduce transport properties from simple scaling arguments. The RG calculation will be done on the replicated partition function $\mathcal{Z}_{\textrm{rep}} \equiv \textrm{Tr} \left[ e^{-\beta \mathcal{H}_{0,\textrm{rep}}} \overline{\hat{U}_{\textrm{rep}}}(\beta,0) \right] $, which is here the quantity of interest.~\cite{Gia}

%
%

\section{Effects of random disorder on interactions and two-particle backscattering}

\subsection{RG flow equations}

\begin{figure}
 \includegraphics[width=0.46\textwidth]{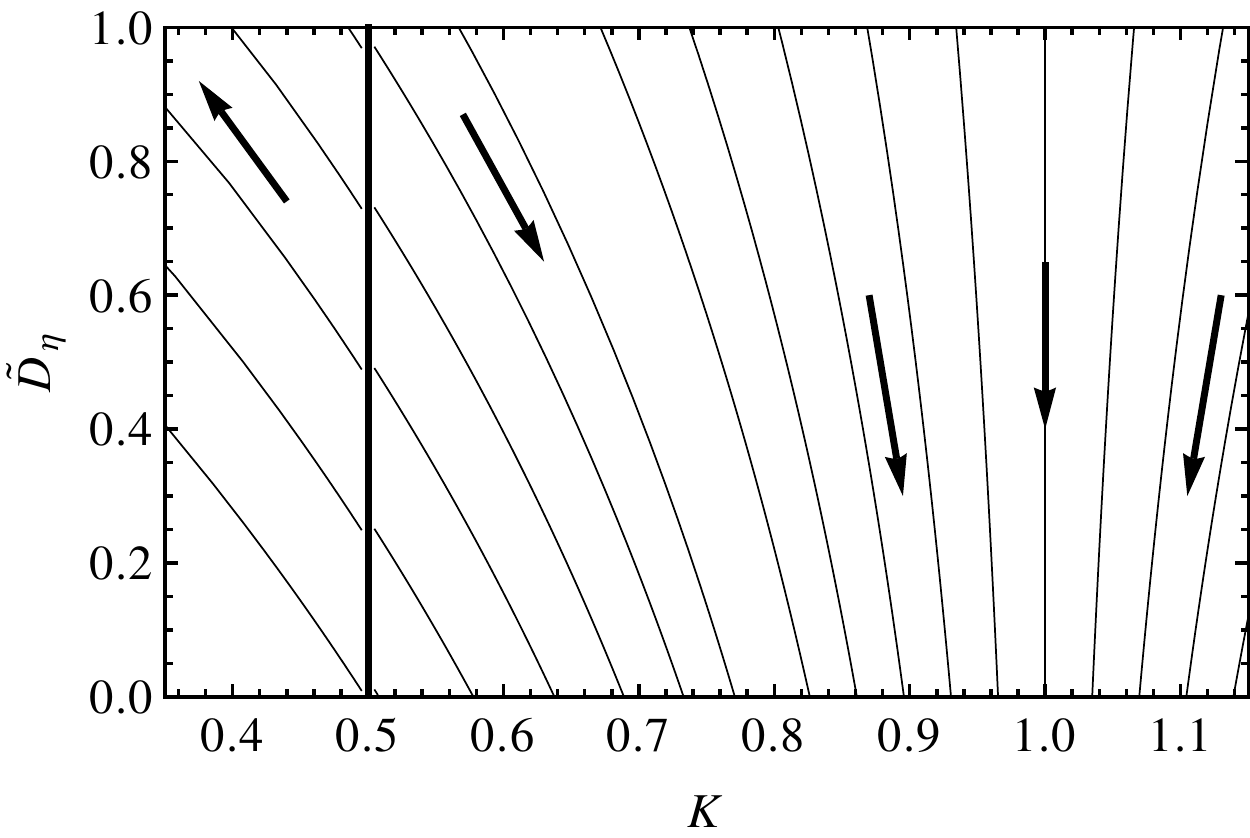}
\caption{RG flow in the plane ($K$,$\DeT$) of parameter space. Disorder becomes relevant for $K<1/2$ and is renormalized to zero otherwise.}
\label{fig:Plot1}
\end{figure}

So far, we have derived an effective Hamiltonian, $\hat{\mathcal{H}}_{\textrm{dis}}$, that takes into account Coulomb interactions exactly and included effects of the disordered Rashba potential through the coupling of replicas. The resulting effective operator in Eq.~(\ref{eq:DefSRrepNew}) therefore represents the full influence of Rashba disorder on an interacting helical electron system. It is in principle able to generate or renormalize different kinds of scattering processes. These contributions arise naturally, at each order in a perturbative expansion of the replicated partition function. This expansion is controlled by the dimensionless Rashba parameter $\DeT=\De/(a v^2)$. We adopt a real space RG scheme~\cite{Car96}, by rescaling the short distance cutoff $a$ to $a'=a(1+d\ell)$, in an OPE.
To first order in $\DeT$, we find that disorder allows for a term of the form
\begin{equation}\label{eq:DFirstOrder}
\frac{2 \DeT v}{\pi^2} (1-K)(1-2K) (1+d\ell)  \int dx d\tau \ \pp (\partial_x \theta(x,\tau))^2 \pp \;, 
\end{equation}
in the expansion.
Details of this calculation are outlined in Appendix \ref{sec:AppA}. After reexponentiation, the product of interaction parameters $Kv$, in the free bosonized Hamiltonian $\mathcal{H}_0$, is renormalized. At the same time, $v/K$ is not, since no contribution proportional to $\pp (\partial_x \phi)^2 \pp$ arises.
Hence, both the interaction parameter $K$ and the plasmon velocity $v$ are renormalized. Moreover, the renormalization of the Rashba disorder parameter $\DeT$ can directly be extracted from Eq.~(\ref{eq:DefSRrepNew}). Putting all results together, we find the following flow equations to first order in $\DeT$
\begin{align}\label{eq:FlowK}
&  \frac{dK}{d\ell}(\ell)=-\frac{2 \DeT(\ell)}{\pi} \left(1-K(\ell)\right)\left(1-2K(\ell)\right), \\
&  \frac{dv}{d\ell}(\ell)=-\frac{2 \DeT(\ell) v(\ell)}{K(\ell)\pi} \left(1-K(\ell)\right)\left(1-2K(\ell)\right), \\
\label{eq:FlowDtilde}
&  \frac{d\DeT}{d\ell} (\ell) = \left(1-2K(\ell)\right) \DeT(\ell).
\end{align}
The resulting flow diagram is shown in Fig.~(\ref{fig:Plot1}). Importantly, no inelastic interactions can be generated by elastic disorder in the non-interacting limit $K\to 1$, as it should be. This correct limit appears in the flow equations even without the additional implementation of a missing piece to cure the fact that we introduced a cutoff on the time variables. The concept of a missing piece, developed in Ref.~\onlinecite{GiaPa} and for instance applied in Ref.~\onlinecite{Crepin1}, is in general needed when working with a real-space RG. Its implementation is crucial to carefully distinguish between elastic and inelastic scattering processes and to obtain correct non-interacting limits. More comments and explanations about our approach to handle this point are given in Appendix \ref{sec:AppA0}. The reason, that no missing piece is needed here for the correct limit, is the following: In first order of $\DeT$, disorder is formally not able to produce an interaction term of the $g_2$ type, because of the derivatives in the Rashba Hamiltonian. The effect of Rashba disorder in
this order of the perturbation can thus be seen only as a renormalization of the effective Fermi velocity. We find that the Rashba disordered potential is an irrelevant perturbation as long as $K>1/2$. In the plane ($K$,$\DeT$), there is a line of fixed points at $K=1/2$ and for $K<1/2$, the system flows to strong coupling, away from the perturbative regime. The question of finding the strong-coupling fixed point and whether it corresponds to Anderson localization remains open.

%

\begin{figure*}
\begin{minipage}[h]{.48\textwidth}
\centering
 \vspace{0.3cm}
\includegraphics[width=0.85\textwidth]{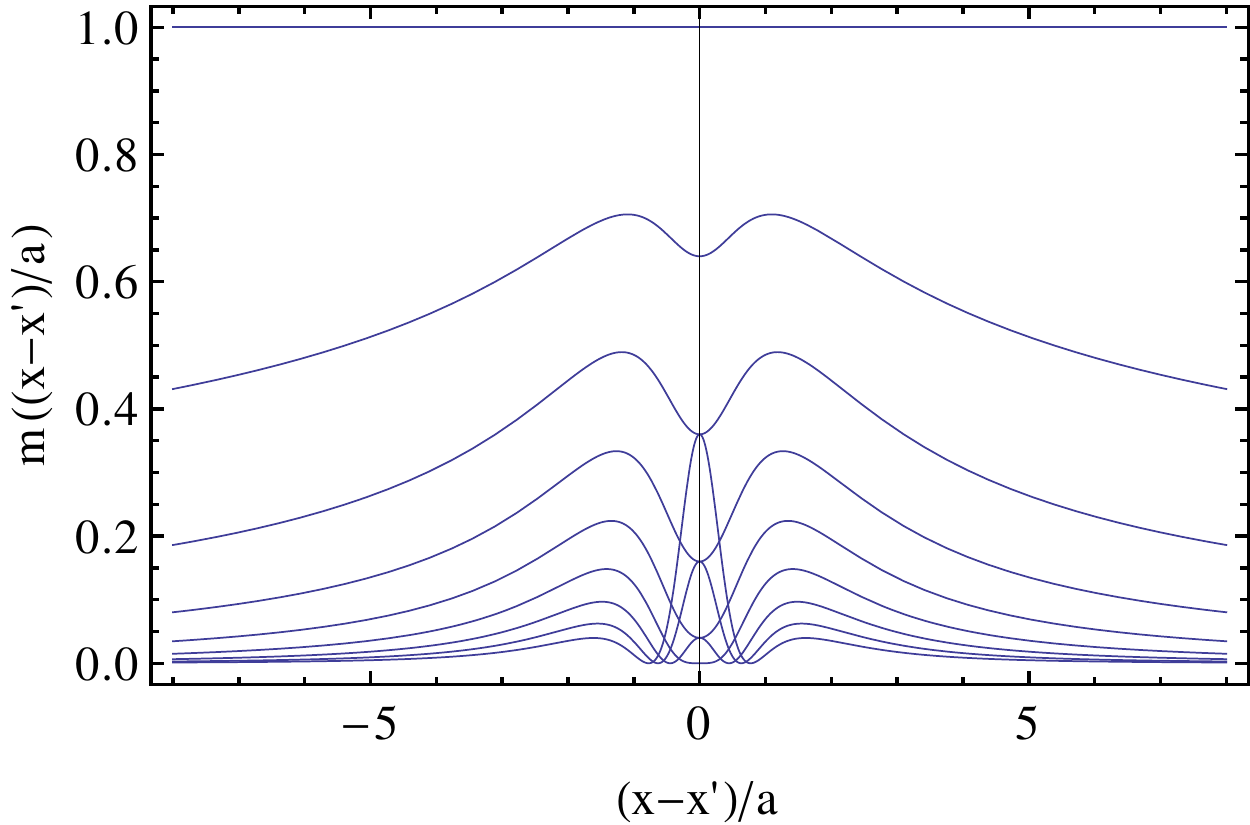} \\
\centering{(a)}
\end{minipage}
\begin{minipage}[h]{.48\textwidth}
\centering
\vspace{0.5cm}
\includegraphics[width=1.\textwidth]{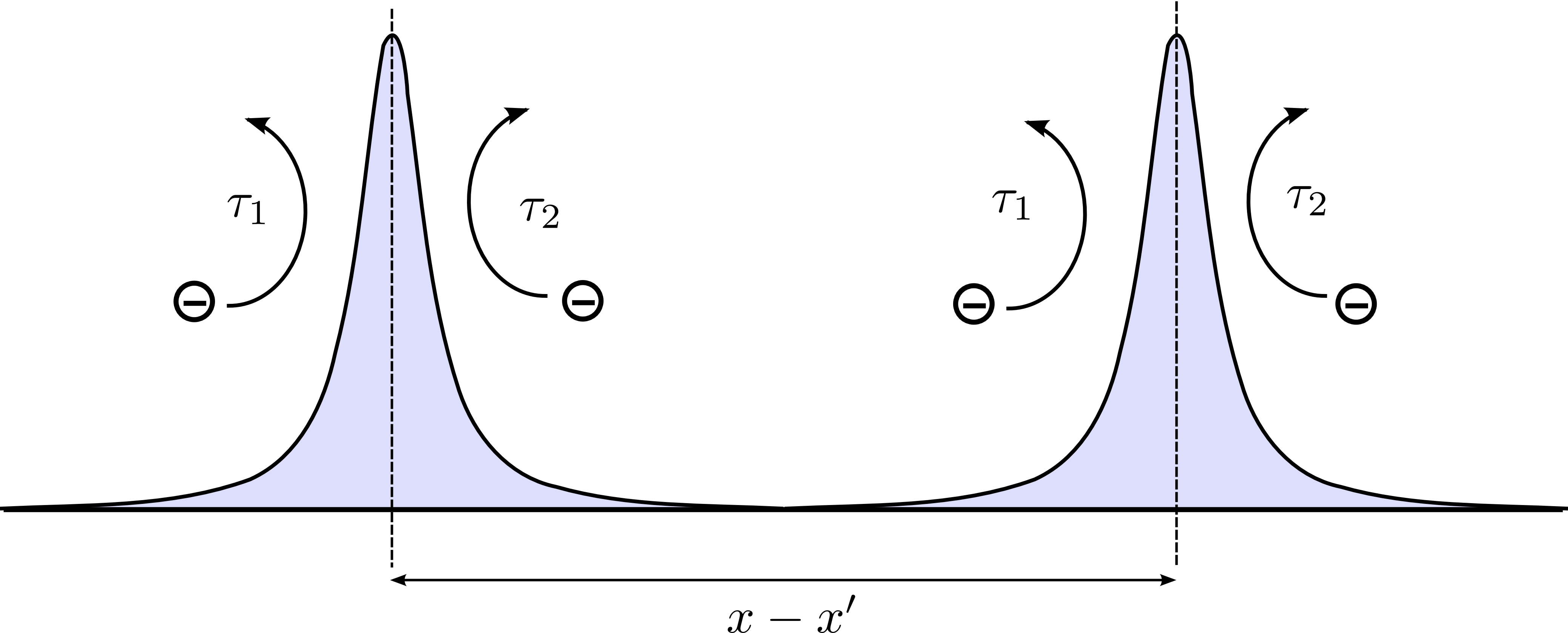} \\
\vspace{0.8cm}
\centering{(b)}
\end{minipage}
\caption{(Color online) (a) Schematic of the form factor $m\left(\frac{x-x'}{a}\right)$ for different values of the interaction parameter $K$, decreasing from top to bottom in steps of $0.1$ starting from $K=1.0$ (top). The dip at $(x-x')=0$ for $K>1/2$, as well 
as the additional maxima away from $x-x'=0$ for $K<1/2$ are a
consequence of the finite cutoff $a$ on our model. (b) Illustration of a TPB-process with two scattering events taking place at a finite spatial distance. Two left-moving particles are backscattered into two right-moving particles and vice versa.}
\label{fig:Plot1b}
\end{figure*}

As explained in the introduction, TRS forbids elastic single-particle backscattering from the Rashba potential. However, inelastic TPB is allowed by symmetry provided Coulomb interactions are present. When the system is not at half-filling, such a process cannot be generated in first order of the disorder strength. Going to second order in the perturbation, we find that, after rescaling of the cutoff, the normal-ordered product contains terms of the form
\begin{widetext}
\begin{align}
(1+2d\ell) \frac{1}{2 a^4} \left(\frac{\DeT v}{\pi^2 K} \right)^2 \left(\frac{2\pi a}{L}\right)^{8K} \sum_{a,b} \int dx dx' d\tau_1 d\tau_2 \  m\left(\frac{x-x'}{a}\right) \pp  e^{i 2 \phi_a(x, \tau_1)}  e^{i 2\phi_a(x', \tau_1)}\pp \pp  e^{-i 2 \phi_b(x, \tau_2)}  e^{-i 2 \phi_b(x', \tau_2)} \pp \label{eq:X22pFinal}.
\end{align}
\end{widetext}
Details of the calculation are given in Appendix \ref{sec:AppB}. Here, the part of the time integral below the cutoff, i.e. the missing piece of the real-space RG treatment, was implemented in a similar way as in Ref.~\onlinecite{Crepin1}.
In the integrand of Eq.~\eqref{eq:X22pFinal}, there appears a space-dependent factor weighting the full expression which we have defined as
\begin{equation}
 m\left(\frac{x-x'}{a}\right)=\left[\frac{(1-2K)-\left(\frac{x-x'}{a}\right)^2}{\left(1+\left(\frac{x-x'}{a}\right)^2\right)^{2-K}}\right]^2.
\end{equation}
It can be verified, for example by going back to fermionic language, that Eq.~(\ref{eq:X22pFinal}) corresponds to a TPB-process, where two left-movers are scattered into two right-movers and vice versa, all in the presence of electron-electron interactions (see Fig.~(\ref{fig:Plot1b}b)). We emphasize, that although each backscattering event was constricted to one spatial point, as $\overline{\eta^*(x_1) \eta(x_2)}=\De \delta(x_1-x_2)$, there is no reason for \textit{two} scattering events to be local in space. 
Such a non-local TPB process is modulated by the form factor $m\left((x-x')/a\right)$, which damps the amplitude for large spatial distances $|x-x'|$ as a power law. The function $m\left((x-x')/a\right)$ is plotted in Fig.~(\ref{fig:Plot1b}a). Note however that the form factor is actually a constant function in the non-interacting case, illustrating the fact that no {\it correlated} two-particle backscattering occurs in the absence of Coulomb interactions.

We now derive the flow equation corresponding to the TPB process generated in Eq.~\eqref{eq:X22pFinal}. Since only the distance of both positions is of importance, let us introduce new coordinates
$\xi=x-x'$ and $\Xi=(x+x')/2$. First, we keep $\xi$ finite, referring to Eq.~(\ref{eq:X22pFinal}) as a \textit{non-local} TPB process. The generated operator is then of the form
\begin{widetext}
\begin{align}
\label{eq:H22pPr}
 \int d\tau_1 d\tau_2 \hat{\mathcal{H}}_{2p} (\tau_1, \tau_2)& =  \gamma_{2p} \left(\frac{2\pi a}{L}\right)^{8K} \frac{v^2}{a^4} \sum_{a,b} \int d\Xi d\xi d\tau_1 d\tau_2\ m\left(\frac{\xi}{a}\right) \notag \\
& \times \pp e^{i 2 \phi_a(\Xi+\frac{\xi}{2}, \tau_1)}  e^{i2 \phi_a(\Xi-\frac{\xi}{2}, \tau_1)} \pp \pp  e^{-i 2 \phi_b(\Xi+\frac{\xi}{2}, \tau_2)}  e^{-i2 \phi_b(\Xi-\frac{\xi}{2}, \tau_2)} \pp .
\end{align}
\end{widetext}
Here, the form factor $m\left(\frac{\xi}{a}\right)$ inside the integral depends on the cutoff. Let us therefore keep $\xi$ as a parameter and analyze the flow equation of the space-dependent TPB-process $\gamma_{2p}(\ell, \xi)$. We are only interested in the inelastic component of these processes, $\gamma_{2p}^{in}(\ell, \xi)$, defined by $\gamma_{2p}(l, \xi)=\gamma_{2p}^{in}(l, \xi)+\frac{\DeT^2}{2 \pi^4 K^2}$. We then arrive at the following flow equation
\begin{align}\label{eq:Nonlocalg}
  \frac{d}{d\ell}\gamma_{2p}^{in}(\ell, \xi)&={\gamma}_{2p}^{in}(\ell, \xi)\left[(4-8K) +\frac{\xi}{a(\ell)} \frac{m'\left(\frac{\xi}{a(\ell)}\right)}{m\left(\frac{\xi}{a(\ell)}\right)} \right] \notag \\
& + \frac{\DeT^2}{2 \pi^4 K^2}\left[(4-4K)+\frac{\xi}{a(\ell)} \frac{m'\left(\frac{\xi}{a(\ell)}\right)}{m\left(\frac{\xi}{a(\ell)}\right)} \right].
\end{align}
Eq.~(\ref{eq:Nonlocalg}) provides, in principle, a full solution for the evolution of $\gamma_{2p}^{in}(l, \xi)$ depending crucially on the ratio of spatial distance between the scattering events and the cutoff.
While $a$ is growing with the RG flow, but is stopped by the lesser of $\xi$ or $\beta$, $\xi$ can take all values up to the length of the system. Let us now illustrate the result on the basis of two limits. In the regime where $\xi \ll a$, one obtains the flow equation
\begin{align}\label{eq:smallXi}
 &  \frac{d}{d\ell}\gamma_{2p}^{in}(\ell, \xi \ll a)\sim {\gamma}_{2p}^{in}(\ell)(4-8K) + \frac{\tilde{\De}(\ell)^2}{2 \pi^4 K^2}(4-4K)\;.
\end{align}
If on the other hand $\xi \gg a$, the flow equation yields
\begin{align}\label{eq:largeXi}
 &  \frac{d}{d\ell}\gamma_{2p}^{in}(\ell, \xi \gg a)\sim - 4K {\gamma}_{2p}^{in}(\ell)+ \frac{12\tilde{\De}(\ell)^2}{2 \pi^4 K^2}\frac{ a(\ell)^2 (1-K)}{\xi^2}\;.
\end{align}
Due to the specific decay of the factor $m$ with large $\xi$, all contributions stemming from the Rashba disorder $\DeT$ are surpressed in this limit. Assuming further that $\gtp(0)=0$, inelastic TPB is never generated at inifinitely large spatial distances. \\

\subsection{Local vs. non-local two-particle backscattering processes}

The function $m(\xi/a)$ that modulates the two-particle backscattering processes is a direct consequence of the peculiar form of the Rashba potential in bosonization. However, one could argue that a different RG scheme could lead to local two-particle backscattering processes only. This would in practice be the case if we were to rescale both time {\it and} position, for instance by imposing an isotropic cutoff in space-time, $x^2 + (v\tau)^2>a$. Up to unimportant prefactors, $m(\xi)$ would then reduce to a Dirac delta function, and the two-particle backscattering operator would recover its expected $3-8K$ scaling dimension, similarly to Ref.~\onlinecite{Stroem}. However, since disorder explicitly breaks Lorentz invariance at the edge --  disorder average only restores translational invariance --  such an isotropic treatment of time and space in the RG is in our opinion not justifiable, at least not without additional assumptions on the microscopic details of the model. Therefore, our treatment helps uncovering a backscattering process that was so far not realized, namely the appearance of non-local two-particle backscattering processes.  In the next section, we discuss the possible signatures of such processes in the edge conductance.

\section{Conductance}
Our analysis of the conductance in this section is twofold. With the help of the Kubo formula, we first compute the finite temperature corrections to the conductance arising from a non-local TPB process of the form of Eq.~(\ref{eq:H22pPr}). 
To that end, we will consider a very large albeit finite wire of length $L$, and verify that the temperature scaling we obtain in the limits of small and large separations $\xi$ is consistent with the RG scaling, if one were to scale the cutoff $a$ from its bare value $a_0 \simeq v/E_G$, with $E_G$ the bulk band gap, up to the thermal length $v\beta$ or to the spatial distance $\xi$, depending on which one is smaller. Note that our calculation relies on the following hierarchy of length scales, $a_0 \ll v\beta \leq L$. With a typical band gap of $E_G \sim 20$~meV, a wire of length $L\sim 5~\mu$m and a plasmon velocity $v\sim 10^{6}$~m.s$^{-1}$, we find that this hierarchy of length scales corresponds to the condition $1.5 ~{\rm K} <  T \ll 200 ~{\rm K}$ on the temperature.  
In the first two subsections of this chapter, our philosophy is to keep $\xi$ as a free parameter, that is restricted only by the system size and the ultraviolet cutoff. 
The physics of the system is then determined by the interplay of $\xi$ and $\beta$, where $a\leq |\xi|\leq L/2$. In fact, we will see that the position-dependent conductance exhibits different scaling behaviors for $\xi\ll v\beta$ and $\xi \gg v\beta$, and we find a qualitative agreement with our RG calculation. In the last subsection, we discuss confinements for $\xi$ respecting finite temperature, and derive a position-independent result for the dc conductance by integrating over all spatial distances. \\

\subsection{Kubo formula for the conductance}
Starting from the Kubo formula, the non-local ac conductivity of the system, at frequency $\omega$, is given by \cite{Maslov2, Gia}
\begin{align}
\label{eq:sigma}
 & \sigma_{\omega}(x,x')=\left.\frac{i e^2 \omega_n^2}{\pi \omega} G_{\omega_n}(x,x')\right|_{\omega_n \to \omega-i\epsilon}, \\
& G_{\omega_n}(x,x')=\int_0^{\beta} d\tau \langle \mathcal{T}_{\tau} \phi(x,\tau) \phi(x',0)\rangle e^{-i \omega_n \tau}\;, \notag
\end{align}
where $ G_{\omega_n}(x,x')$ is the Fourier transform of the imaginary time boson propagator, and $\omega_n$ a Matsubara frequency. Implementing the TPB Hamiltonian on the basis of Eq.~(\ref{eq:H22pPr}), we compute the correction $\delta G_{\omega_n}(x,x')$ to the free Green's function, to first order in $\hat{\mathcal{H}}_{2p}$. Following Eq.~\eqref{Eq:corr2}, we obtain the following expression
\begin{widetext}
\begin{align}\label{eq:dG2}
\delta G_{\omega_n}(x,x') &= \lim_{N \to 0} \frac{1}{N} \sum_{a,b,c=1}^N
\frac{\gt}{a^4} \int_0^{\beta}d\tau e^{-i \omega_n \tau} \int d\Xi d\xi d\tau_1 d\tau_2\ \mt \left(\frac{\xi}{a}\right) \times \nonumber \\ \langle &\mathcal{T}_{\tau} \phi^a(x,\tau) \phi^a(x',0)
e^{i 2 \phi_b(\Xi+\frac{\xi}{2}, \tau_1)}  e^{i2 \phi_b(\Xi-\frac{\xi}{2}, \tau_1)}
e^{-i 2 \phi_c(\Xi+\frac{\xi}{2}, \tau_2)}  e^{-i2 \phi_c(\Xi-\frac{\xi}{2}, \tau_2)}\rangle_0 \, .
\end{align}
Since averages are performed with respect to the free theory, only terms that are diagonal in replicas survive. Moreover, the replica limit, $N \to 0$, make disconnected diagrams vanish. Note that we have introduced a modified form factor
\begin{equation}
 \mt \left(\frac{\xi}{a}\right)=\left(\frac{(1-2K)-\left(\frac{\xi}{a}\right)^2}{\left(1+\left(\frac{\xi}{a}\right)^2\right)^{2}}\right)^2\;.
\end{equation}
After Fourier transformation, the propagator takes the compact form~\cite{Maslov2} (for ease of notation, we use $x_1=\Xi+\frac{\xi}{2}$ and $x_2=\Xi-\frac{\xi}{2}$)
\begin{align}\label{eq:dG3}
\delta G_{\omega_n}(x,x')& = - 4
\frac{\gt}{a^4} \int dx_1 dx_2 \left[ F_{\omega_n=0}(x_1-x_2)-F_{\omega_n}(x_1-x_2)\right] \mt \left(\frac{x_1-x_2}{a}\right) \notag \\
& \times \left[G^0_{\omega_n}(x,x_1) G^0_{\omega_n}(x',x_1)+ G^0_{\omega_n}(x,x_2) G^0_{\omega_n}(x',x_2)+G^0_{\omega_n}(x,x_1) G^0_{\omega_n}(x',x_2)+G^0_{\omega_n}(x,x_2) G^0_{\omega_n}(x',x_1)\right]\;,
\end{align}
with $F_{\omega_n}$ the Fourier transform of 
\begin{align*}
F(x_1,x_2, \tau_1,\tau_2)= \langle e^{2 i \left( \phi(x_1, \tau_1)+\phi(x_2, \tau_1)- \phi(x_1, \tau_2)- \phi(x_2, \tau_2)\right)} \rangle_0\;.
\end{align*}
Evaluating the free propagator $G^0$, it can be readily seen that in the limit of large system sizes $L$, $G^0_{\omega_n}\approx \frac{K}{2 \omega_n}$ \cite{Maslov2}, and is therefore independent of both positions and the inverse temperature. Thus, the temperature correction to the conductance originates exclusively from the function $F_{\omega_n}(x_1-x_2)$. We write~\cite{Gia} $F(x_1,x_2, \tau_1, \tau_2)= \exp[-2 [F_1(x_1-x_2,0) - F_1(0, \tau_1-\tau_2) - F_1(x_1-x_2, \tau_1-\tau_2) -F_1(x_1-x_2, \tau_1-\tau_2)-F_1(0, \tau_1-\tau_2)+F_1(x_1-x_2,0)]$, with
\begin{equation*}
F_1(x_1-x_2, \tau_1-\tau_2)= K\langle [\phi(x_1,\tau_1)-\phi(x_2,\tau_2)]^2\rangle_0.
\end{equation*}
The correlation function $F_1$ can be derived along the lines of Refs. \onlinecite{Gia, Delft}. For finite temperature and large system sizes, returning to the notation $\xi=x_1-x_2$,
\begin{align}\label{eq:F1}
 F_1(\xi, \tau_1,\tau_2)=\frac{K}{4}\log\left[ \frac{ \left( \sinh^2\left(\frac{\pi \xi}{v\beta}\right) +\sin^2\left(\frac{\pi}{v\beta} (v\tau_1-v\tau_2-a)\right) \right)
\left( \sinh^2\left(\frac{\pi \xi}{v\beta}\right) +\sin^2\left(\frac{\pi}{v\beta} (v\tau_1-v\tau_2+a)\right) \right)
 }{\sin^4\left(\frac{\pi a}{v\beta}\right)} \right]\;.
\end{align}
\end{widetext}
Using Eq.~(\ref{eq:F1}), a general expression for $F(\xi, \tau)$ can be derived, although an exact Fourier transformation to frequency space remains difficult. In the two limits $\xi\to 0$ and $\xi \to  \infty$, we obtain the analytical expressions
\begin{align*}
& F(\xi \to 0, \tau)= \left(\frac{\pi a}{v\beta}\right)^{8 K} \left(\sin\left(\frac{\pi \tau}{\beta}\right)\right)^{-8K}\;, \\
& F(\xi\to \infty, \tau)= \left(\frac{\pi a}{v\beta}\right)^{4 K} \left(\sin\left(\frac{\pi \tau}{\beta}\right)\right)^{-4K}\;.
\end{align*}
and their Fourier transforms as
\begin{align}
\label{eq:Fw1}
& \left[ F_{\omega_n=0}(\xi)-F_{\omega_n}(\xi)\right]_{\xi \to 0}= C_0 \left(\frac{\pi a}{v\beta}\right)^{8 K} \beta^2 \omega_n  +\mathcal{O}(\omega_n^2)\;, \\
\label{eq:Fw2}
& \left[ F_{\omega_n=0}(\xi)-F_{\omega_n}(\xi)\right]_{\xi\to \infty}= C_{\infty} \left(\frac{\pi a}{v\beta}\right)^{4 K} \beta^2 \omega_n  +\mathcal{O}(\omega_n^2)\;.
\end{align}
Here, $C_0$ and $C_{\infty}$ are geometric factors depending on the parameter $K$. In the dc limit $\omega\to0$, we can then find the temperature dependent correction $\delta g$ to the dc conductance, in the two limits of small and large $\xi$. Using that $\mt\left({\frac{\xi}{a}\to 0}\right)= (1-2K)^2$ and $\mt\left({\frac{\xi}{a}\to \infty}\right)\sim \left(\frac{\xi}{a}\right)^{-4}$, these corrections become
\begin{align}
\label{eq:gfinalsmallxi}
 & \delta g(\xi \to 0)= -\frac{e^2}{\pi}4 L C_0 K^2(1-2K)^2\ \gt a^{-4} \left(\frac{v\beta}{\pi a}\right)^{-8K} \beta^{2}, \\
\label{eq:gfinallargexi}
 & \delta g(\xi \to \infty) = -\frac{e^2}{\pi} 4 L C_{\infty} K^2\ \gt\ \xi^{-4} \left(\frac{v\beta}{\pi a}\right)^{-4K} \beta^{2}.
\end{align}
We find, that the correction to the conductance scales as $\delta g\sim \beta^{2-8K}$ for $\xi \to 0$ and as $\delta g\sim \beta^{2-4K}$ for $\xi \to \infty$, if it is exclusively dominated  by the parameter $\gt$.
Let us now compare these results with the RG flow equations~\eqref{eq:smallXi} and ~\eqref{eq:largeXi}. As a first approximation we neglect the contribution of $\tilde{D}_\eta$. A simple consistency check is then to take $a = v\beta$ in Eqs.~\eqref{eq:gfinalsmallxi} and \eqref{eq:gfinallargexi} {\it and} rescale $\gt$ accordingly, following either Eq.~\eqref{eq:smallXi}, $\gt \sim (v\beta/a_0)^{4-8K}$, or Eq.~\eqref{eq:largeXi}, $\gt \sim (v\beta/a_0)^{-4K}$. We recover in both cases the same temperature scaling, a result that strengthens our RG approach for the non-local two-particle processes.

\subsection{RG approach to the conductance}

\begin{figure}[Htb]
\includegraphics[width=0.45\textwidth]{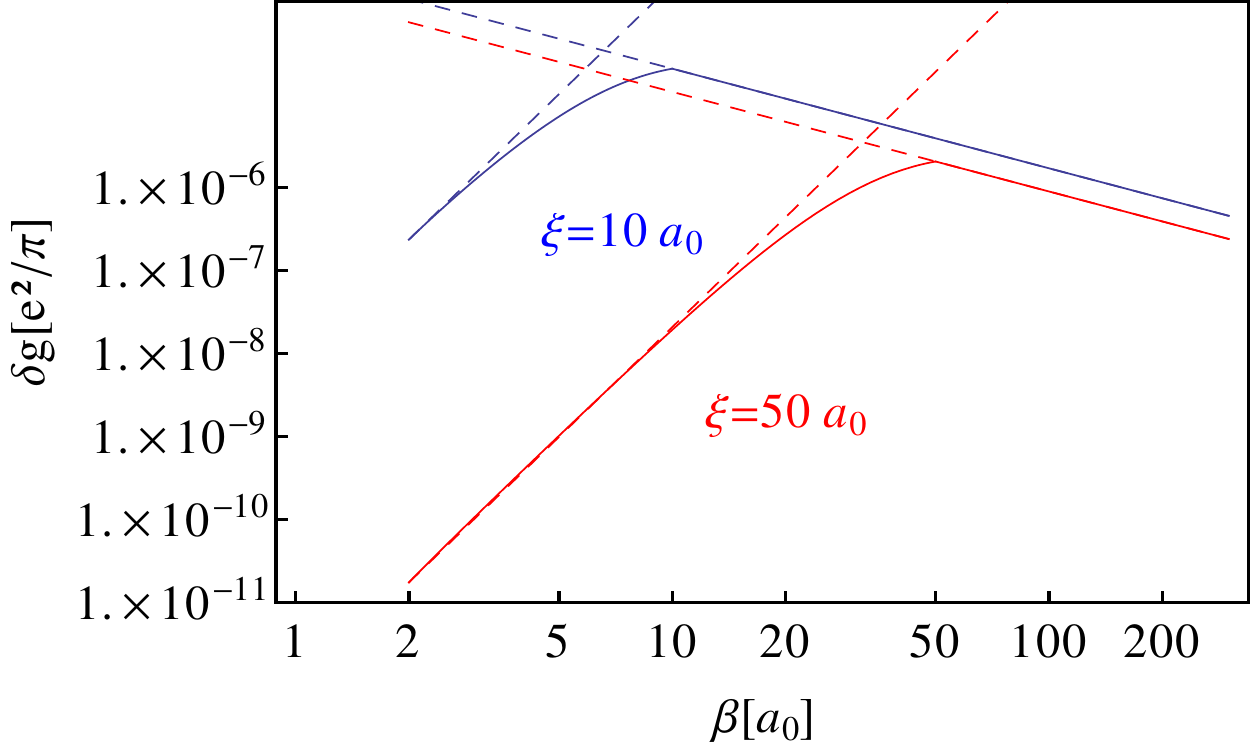}
\caption{(Color online) Plot of the correction to the conductance $\delta g$ depending on inverse temperature $\beta$. 
The spatial separation $\xi$ is kept fixed, and two exemplary values are shown to illustrate the damping of the conductance for increasing distances. 
The plot is composed of two parts, where in the regime $\xi<\beta$ and $\xi>\beta$ the flow of the cutoff parameter $a$ was stopped at $\xi$ or $\beta$, respectively, and the two scales cross over at $\xi \sim \beta$. Note, that for $\xi < \beta$ we find 
a perfect powerlaw, since the solution for $\gt$ does not depend on $\beta$. The powerlaw dependencies of $\delta g$ in the limit of small and large distances, as given in Eqs.~(\ref{eq:PowerLaw1}) and ~(\ref{eq:PowerLaw2}), are visualized as dashed lines.  
We have chosen parameters $\DeT(0)=0.01, \gtp(a_0)=0, a_0=1, v=1, L=300 a_0$, as well as $K=0.4$.}
\label{fig:Plot2}
\end{figure}


%

In this section we refine our analysis of the conductance corrections. Since in our theory, two-particle backscattering processes are actually generated by the Rashba disordered potential, it is necessary to consider the full set of RG flow equations. We combine our calculation of the conductance using the Kubo formula with the solution of the flow equations \eqref{eq:FlowK},~\eqref{eq:FlowDtilde},~\eqref{eq:Nonlocalg}. In this approach, the corrections to the conductance, for a given separation $\xi \ll \beta $, are given by, 
\begin{align}\label{eq:NonlocalG}
  \delta g(\xi, \beta) \simeq -\frac{e^2}{\pi} 4 L  K^2 \beta^{2} \left( \frac{\pi a(\ell)}{v\beta} \right)^{8K} \frac{\gtp(\ell,\xi)}{a(\ell)^4}  \mt\left(\frac{\xi}{a(\ell)}\right)\;,
\end{align}
with $\gtp(\ell,\xi)$ the solution of Eq.~\eqref{eq:Nonlocalg}. 
Motivated by our analysis of Eqs.~\eqref{eq:gfinalsmallxi} and \eqref{eq:gfinallargexi},  we have assumed that the $\xi$ dependence is well captured by both $\gtp(\ell,\xi)$ and $\mt(\xi/a)$, and that the function $F$ of Eq.~\eqref{eq:dG3} only contributes unimportant geometric factors. With the boundary conditions $\gtp(\ell=0,\xi)=0$, $\DeT(\ell=0) = \DeT^0 $, and keeping $K$ constant, we arrive at the following expressions, in the limiting cases, 
\begin{align}
\gtp(\ell,\xi) \simeq \frac{2(\DeT^0)^2}{\pi^4 K^2}\frac{K-1}{2K-1}\left(e^{(4-8K)\ell}-e^{(2-4K)\ell} \right)\;.
\end{align}
The order of length scales being $a_0 \ll \xi \ll \beta$, the flow should be stopped at $a(\ell) = \xi$, or equivalently $\ell = \ln (\xi/a_0)$. We then find two different scalings crossing over at $K=1/2$. Indeed, if $1/2<K<1$, the scaling is dominated by $\DeT$, leading to $\gtp(\beta, \xi )\sim \xi^{2-4K}$, while if $K<1/2$, we find instead $\gtp(\beta, \xi)\sim \xi^{4-8K}$. Using $\mt \left(\xi \to 0\right) \to (1-2K)^2$ in Eq.~\eqref{eq:NonlocalG}, as well as $a(\ell) = \xi$, we finally arrive at 
\begin{align}\label{eq:PowerLaw1}
&\delta g ( \xi \ll v\beta ) \sim \begin{cases}
		 -(v \beta)^{2-8K} \xi^{4K-2}  &\mbox{if } 1/2<K<1\;, \\
                 -(v \beta)^{2-8 K}  &\mbox{if } K<1/2\;.
            \end{cases}
\end{align}
In the opposite regime, $a_0 \ll \beta \ll \xi$, the flow equation for $\gtp(\ell,\xi)$ gives
\begin{align}
\gtp(\ell,\xi) \simeq \frac{6(\DeT^0)^2}{\pi^4 K^2}\frac{K-1}{K}\frac{a_0^2}{\xi^2}\left(e^{-4K\ell}-e^{(4-4K)\ell} \right)\;.
\end{align}
The flow should in this case be stopped at $a(\ell) = \beta$, or equivalently $\ell = \ln (\beta/a_0)$. The scaling is always dominated by $\DeT$ and $\gtp(\beta, \xi\to \infty) \sim \beta^{4-4K}\xi^{-2}$. Using now $\mt \left(\xi \to \infty\right)\sim \left(\frac{\xi}{\beta}\right)^{-4}$, we obtain
\begin{align}\label{eq:PowerLaw2}
&\delta g ( \xi \gg v\beta ) \sim -(v \beta)^{6-4 K} \xi^{-6}\;.
\end{align}
 The scaling behavior of the conductance with temperature is illustrated in Fig.~(\ref{fig:Plot2}). Different power laws are found for small and large spatial distances. However,
it is important to remember that besides the scaling, the full expression is damped by the factor $\mt$ with increasing $\xi$.
We find that the scaling laws of $\delta g(\xi,\beta)$ are very robust against perturbations of $\xi$ or $\beta$, as long as the
ratio of $\xi/\beta$ is not changed dramatically. If thermal and spatial lengths become of the same order, there is a crossover between two scalings. 

\subsection{ Position-independent conductance}


In this last subsection, we would like to sketch a way how to derive a result for the correction to the conductance that is position-independent. We come back to a more microscopic analysis in order to integrate over the space separation $\xi$, respecting the hierarchy of length scales $a_0 \ll \xi \ll \beta$. 
Here, the cutoff does not flow, so we fix $a=a_0$.
The important point to note is the contraction of two out of four times along the normal-ordering process (see Eq.~(\ref{eq:FullProduct2})), that we performed for simplicity. 
Choosing the same two-particle backscattering term, Eq.~(\ref{eq:H22pPr}) before the contraction of time variables reads 
\begin{widetext}
 \begin{align}
\label{eq:H22pPrNew}
\int d\tau_1 d\tau_2 d\tau_3 d\tau_4 \hat{\mathcal{H}}_{2p} (\tau_1, \tau_2, \tau_3, \tau_4)& = 
\gamma_{2p} \left(\frac{2\pi a_0}{L}\right)^{8K} \frac{1}{a_0^4} \sum_{a,b} \int d\Xi dY dY'\ \int d\xi\ m_2(\xi,\beta) \notag \\
& \times \pp e^{i 2 \phi_a(\Xi+\frac{\xi}{2}, \tau_1)}  e^{i2 \phi_a(\Xi-\frac{\xi}{2}, \tau_1)} \pp \pp  e^{-i 2 \phi_b(\Xi+\frac{\xi}{2}, \tau_2)}  e^{-i2 \phi_b(\Xi-\frac{\xi}{2}, \tau_2)} \pp .
\end{align}
where $Y=\frac{1}{2}(y_1+y_3)$, $y=y_1-y_3$, $Y'=\frac{1}{2}(y_2+y_4)$ and $y'=y_2-y_4$. Here, we have defined an effective form factor of 

\begin{align}\label{eq:m2}
 m_2(\xi,\beta)&=a_0^{-4K+2} \int_0^{v\beta} dy \frac{1}{4}\frac{(1-2K)(y+a_0)^2-\xi^2}{((y+a_0)^2+\xi^2)^{2-K}}\ \int_0^{v\beta} dy'\frac{(1-2K)(y'+a_0)^2-\xi^2}{((y'+a_0)^2+\xi^2)^{2-K}}\simeq \notag \\
& a_0^{-4K+2}[a_0^2 \xi^{4K-4}-2a_0\xi^{2K-2}(v\beta)^{2K-1}+(v\beta)^{4K-2}].
\end{align}
\end{widetext}
Here, we used again $a_0 \ll \xi \ll \beta$.
Next, we expand the exponentials between normal-ordering signs in Eq.~(\ref{eq:H22pPrNew}) in terms of $\xi$, and use again Eq.~(\ref{eq:Fw1}) to calculate the correction to the conductance. Similar to Eq.~(\ref{eq:NonlocalG}), we have
\begin{align}\label{eq:dgnew}
 \delta g(\xi, \beta) \sim -\frac{\gamma_{2p}}{a_0^4} \beta^2 \left(\frac{\pi a_0}{v\beta}\right)^{8K} m_2(\xi,\beta).
\end{align}
By inspection of Eq.~(\ref{eq:m2}), we find that for $K<1/2$, all terms depending on temperature are very small, and the leading contribution is $m_2\simeq \left(\frac{\xi}{a_0}\right)^{4K-4}$. On the other hand, if 
$1/2<K<1$, the factor $\beta$ dominates over $a_0$ and we find $m_2 \simeq \left(\frac{v\beta}{a_0}\right)^{4K-2}$. This leads us to 
\begin{align}\label{eq:PowerLaw1New}
&\delta g (\xi) \sim \begin{cases}
		 -(v \beta)^{-4 K} \gamma_{2p} a_0^{4K-2} &\mbox{if } 1/2<K<1\;, \\
                 -(v \beta)^{2-8 K}  \gamma_{2p}\ \xi^{4K-4} a_0^{4K} &\mbox{if } K<1/2\;.
            \end{cases}
\end{align}
Importantly, $\gamma_{2p}$ is here a temperature-independent parameter. Finally, we perform the integration over $\xi$. Since the exponentials in Eq.~(\ref{eq:H22pPrNew}) were expanded to lowest order in $\xi$, the only term depending on $\xi$ in this approximation is $m_2$, and we find
\begin{align}
\int_{a_0}^{v \beta} d\xi m_2(\xi, \beta)&= a_0^{-4K+2}[\beta^{4K-1} -\frac{1}{4K-3}a_0^{4K-1} \notag \\
&+\mathcal{O}(\beta^{4K-2}, \beta^{2K-1}, \beta^{4K-3})]. \notag
\end{align}
The terms in the inner brackets are all subleading in $\beta$ and can therefore be disregarded. We note, that for $K<1/4$, the dominant contribution is again the term depending on 
the cutoff $\int d\xi m_2(\xi) \simeq a_0$. If on the contrary $1/4<K<1$, the leading term is $\int d\xi m_2(\xi) \simeq a_0\left(\frac{v\beta}{a_0}\right)^{4K-1}$. 
 With the help of Eq.~(\ref{eq:dgnew}), we conclude, that the position-independent 
corrections to the dc conductance $\delta G \simeq \int_{a_0}^{v\beta} d\xi\ \delta g ( \xi)$ scale as
\begin{align}\label{eq:PowerLaw2new}
\delta G \sim \begin{cases}
		 -(v \beta)^{1-4 K} \gamma_{2p} a_0^{4K-2} &\mbox{if } 1/4<K<1\;, \\
                 -(v \beta)^{2-8 K}  \gamma_{2p}\ a_0^{8K-3} &\mbox{if } K<1/4\;.
            \end{cases}
\end{align}
Due to the integration over $\xi$, the crossover of scalings was shifted to the point $K=1/4$.
With $\beta = 1/(k_B T)$, we arrive at the final result given in Eq.~(\ref{eq:Tscaling}).

\section{Conclusion}
In this article, we have modeled the influence of random Rashba spin orbit coupling on a helical 1D quantum system such as a QSH edge state. It has been shown that inelastic two-particle backscattering
may reduce the electronic transport properties in the presence of time-reversal symmetry. The correction to the conductance scales as a power law with the temperature, where the exponent is determined only by the electron-electron interaction strength $K$.
This tendency could in principle be experimentally observed in a transport measurement at low temperatures.
In our calculation, it was pointed out that the correction to the conductance due to random Rashba spin-orbit coupling is obtained in second order of the disorder strength. This result was achieved in 
the limit of zero voltage bias, but finite temperature.
Furthermore, we analyzed the relevance of the backscattering operators in the RG sense.
Considering two-particle backscattering, its operator is found to become relevant for strong interactions below $K=1/2$, in 
the disordered case, which is in contrast to the threshold of $K=1/4$ in the single impurity case \cite{Crepin1}. 
Moreover, the disorder strength itself is a relevant parameter of the system in the presence of strong interactions, while the single impurity potential remains irrelevant at all interaction strengths. 
A special emphasis has been put on the possibility of a non-local two-particle backscattering, which is a characteristic feature of the disordered system. We have shown that all such processes
naturally contribute to the conductance correction, though being damped as a power law for large distances. 
Concerning the temperature dependence of the conductance, we find different
scalings depending on the ratio of the spatial distance and the thermal length. The dominant scalings, however, are provided by the regime of spatially coinciding scattering events. 
Integrating out the spatial distance between both events, the correction to the conductance for $1/4<K<1$ is reduced by one power of the temperature compared to the case of two-particle backscattering off a single impurity \cite{Crepin1}.\\ \\
We thank F. Dolcini, I.V. Gornyi, N. Kainaris, A. Mirlin and D. Polyakov for helpful discussions.
Financial support by the DFG (German-Japanese research unit ``Topotronics'' and the priority program SPP 1666 ``Topological insulators''), as well as the Helmholtz Foundation (VITI) is gratefully acknowledged.



\appendix
\onecolumngrid
\section{First order expansion}
\label{sec:AppA}
In this Appendix, we give some details about the procedure of normal-ordering. Let us first consider the expansion of the partition function in first order of $\DeT$, leading to the term given in Eq.~(\ref{eq:DFirstOrder}).

We use $y=v\tau$ as new time variables and further exploit the fact that right-moving fields only depend on the combination $z=ix+y$ and left-moving fields on $\zo=-ix+y$ to
shorten the notation. Moreover, in $\mathcal{H}_0$,  we rescale the fields as $\sk \theta \to \theta$ and $\phi /\sk \to \phi$. Also, we abbreviate generally $\lambda=2\sk$. For fields like $\phi$ and $\theta$, that include both right and left movers, we simply write $\phi(x_1,y_1)=\phi(z_1, \zo_1)=\phi(1)$.\\
%
To normal-order the operator product we use the known commutation relations in the limit of large system sizes $L$, as e.g. given in Ref. \onlinecite{Delft}. Employing general coefficients $\lambda, \lambda'$, we find for the first order term
\begin{align}\label{eq:FullProd1}
&   \pp \partial_x \theta_a(1)   e^{i\lambda \phi_a(1)}  \pp \times \pp \partial_x \theta_b(2)   e^{i\lambda' \phi_b(2)} \pp= h(1,2)\bigg[ \pp \partial_x \theta_a(1) \partial_x \theta_b(2)   e^{i\lambda \phi_a(1)}  e^{i\lambda' \phi_b(2)}  \pp- \notag \\
&  \frac{1}{2} u(1,2)  \pp \left(\frac{\lambda}{2} \partial_x \theta_a(1)- \frac{\lambda'}{2} \partial_x \theta_b(2)\right)   e^{i\lambda \phi_a(1)}   e^{i\lambda' \phi_b(2)}  \pp+ \frac{1}{2^2} s_0(1,2) \pp  e^{i\lambda \phi_a(1)}  e^{i\lambda' \phi_a(2)} \pp \bigg]
\end{align}
with the functions
\begin{align}
&  h(1,2)= \left(\frac{2\pi}{L} |z_1-z_2+a| \right)^{\frac{\lambda \lambda'}{2}}, \notag \\
&  u(1,2)=\frac{1}{z_1-z_2+a}+\frac{1}{\zo_1-\zo_2+a}= \frac{2(y_1-y_2+a)}{|z_1-z_2+a|^2}, \notag \\
&  u_2(1,2)= \frac{1}{(z_1-z_2+a)^2}+ \frac{1}{(\zo_1-\zo_2+a)^2}=\frac{2\left((y_1-y_2+a)^2-(x_1-x_2)^2 \right)}{|z_1-z_2+a|^4}, \notag\\
&  s_0(1,2)=u_2(1,2)-\frac{\lambda \lambda'}{4}(u(1,2))^2=\frac{2\left((1-\frac{\lambda \lambda'}{2})(y_1-y_2+a)^2-(x_1-x_2)^2 \right)}{|z_1-z_2+a|^4}. \notag
\end{align}
Away from half filling, only opposite signs $\lambda'=-\lambda$ are allowed due to the disorder-averaging procedure.
Thus, the first order product can not generate any TPB and we exploit Eq.~(\ref{eq:FullProd1})
just for the renormalization of the Luttinger parameters $K$ and $v$. Switching to coordinates $y= y_1-y_2$ and $Y=(y_1+y_2)/2$, we expand the exponentials in normal-ordering signs around small time distances $y\sim 0$. With the help of the equation of motion $ \partial_Y \phi(x,Y)=-i \partial_x \theta(x,Y)$, we find from Eq.~(\ref{eq:FullProd1})
\begin{align}\label{eq:FullProd1Kv}
& \pp \partial_x \theta(1)   e^{i\lambda \phi(1)}  \pp \times \pp \partial_x \theta(2)   e^{-i\lambda \phi(2)}  \pp \sim h(1,2) \pp (\partial_x \theta(x,Y))^2 \pp \times \notag \\
& \bigg[1- \frac{1}{2} u(1,2) \lambda^2 (y+a)+\frac{1}{2^2} s_0(1,2)\frac{\lambda^2}{2}\ (y+a)^2+ \mathcal{O}(y^3) \bigg].
\end{align}
Plugging the normal-ordered product into Eq.~(\ref{eq:DefSRrepNew}) and remembering that $x_1=x_2=x$, we find
\begin{align}\label{eq:Illustrdl}
&  \exp \bigg[\int d\tau_1 d\tau_2\ \hat{\mathcal{H}}_{dis}(\tau_1, \tau_2)\bigg] \sim  \frac{\DeT}{2} \left(\frac{1}{\pi^2 a K}\right) \left(\frac{2\pi a}{L}\right)^{2K} \int dx dY \ \pp (\partial_x \theta(x,Y))^2 \pp  \\
&  \int dy \left(\frac{2\pi}{L} |y+a| \right)^{\frac{-\lambda^2}{2}} \bigg[1- \lambda^2 +\frac{\lambda^2}{2^2}\ \left(1+\frac{\lambda^2}{2}\right)  \bigg]+H.c.= \notag \\
&   \frac{4 \DeT}{2} \left(\frac{1}{\pi^2 K}\right) (1-K)(1-2K) (1+d\ell+\mathcal{O}(d\ell^2)) \int dx dY \ \pp (\partial_x \theta(x,Y))^2 \pp \notag .
\end{align}
In the last step, we integrated out $y$, putting it to zero on a shell of $2a (1+d\ell)$. As usual we neglected all contributions of order $\mathcal{O}(d\ell^2)$.
After reexponentiating, the Rashba disorder contributes  to the renormalization of the product $Kv$ in $\mathcal{H}_0$ (note that the fields have  to be transformed back $\theta \to \sk \theta$),
\begin{align}\label{eq:FlowKv}
& -\frac{Kv(a')}{2\pi}=-\frac{Kv(a)}{2\pi}+\frac{4 \DeT}{2 \pi^2} (1-K)(1-2K)v\ d\ell, \notag \\
&  \frac{d}{d\ell}(K v)(l)=-\frac{4 \DeT}{\pi} (1-K)(1-2K)v.
\end{align}
Since $v/K$ is not renormalized, the result can be rewritten in terms of $K$ and $v$ separately, as done in Eq.~(\ref{eq:FlowK}).
%
\section{Contractions and missing pieces}
\label{sec:AppA0}
A few comments are in order about the factor $(1+d\ell)$ in Eq.~(\ref{eq:Illustrdl}), where the integer one represents the so-called missing piece now. 
To clarify this point, let us first consider a contraction of time variables in the absence of any cutoff. 
Imagine, that we contract the times $y_1$ and $y_3$ in the following integral
\begin{align}\label{eq:NoMp}
 &\int dy_1 \int dy_3 f(y_1, y_3)= \int dY \int_{-\infty}^{\infty} dy f(Y,|y|)\sim 2\ d\ell \int dY f(Y, 0)= 2\ d\ell \int dy_1 f(y_1, y_1),
\end{align}
where $f$ describes a general function.
In Eq.~(\ref{eq:NoMp}), we have first changed variables to $Y=(y_1+y_3)/2,\ y=y_1-y_3$ and changed $Y$ back to $y_1$ again.
Time-ordering ensures that all time-differences are positive, allowing for the notation of the modulus $|y|$. Assuming that $y_1$ and $y_3$ are very close to each other, a
Taylor expansion can be performed around $y \sim 0$, setting $y$ equal zero on an infinitesimal shell of
size $d\ell$. Proceeding this way, multiple contractions bring factors of $2\ d\ell$ for each variable to be integrated out.\\
Next, we consider the same integral assuming a finite cutoff $a$ on both time variables. The cutoffs are introduced during the bosonization process to avoid divergences, and eventually 
become manifest in the form of replacements $y\to y+a$. The two time-variables to be contracted are then located close to each other on a ring of inner radius $a$ and width $d\ell$. As a first consequence, 
contractions now generate factors of $2a\ d\ell$. 
Second, and more importantly, we miss in any of the time integrals the part $0<y<a$, as was first realized in Ref.~\onlinecite{GiaPa}. This is coined the {\it missing piece} of the RG procedure. At this point, elastic and inelastic characters of scattering processes get mixed, since a purely elastic process would correspond to an unlimited integral.
The missing piece can be implemented together with a rescaling of the cutoff in the following way
\begin{align*}
 & \int_{-\infty}^{\infty} dy f(|y+a|)=2 \int_a^{\infty} dy f(y)\to 2\left(\int_a^{\infty} dy f(y) + Mp \right)= 2\left(\int_a^{a'} dy f(y)+ Mp+\int_{a'}^{\infty} dy f(y)\right)\sim \\
& f(a)\ 2a\ d\ell +2 Mp.
\end{align*}
Here, $(Mp)$ represents the missing piece integral $\int_0^a dy f(y)$. \\
In the spirit of Ref.~\onlinecite{Crepin1}, we obtain
\begin{equation}
 (Mp)= \int_0^a dy f(y)\sim f(a)\ a,
\end{equation}
where a potential divergence at $y=0$ has to be taken with care.
The contraction in Eq.~(\ref{eq:NoMp}), now becomes
\begin{align*}
 &\int dy_1 \int dy_3 f(y_1, y_3)=\int dY \int_{-\infty}^{\infty} dy f(Y,|y+a|) \sim 2a\ (1+d\ell) \int dy_1 f(y_1, y_1).
\end{align*}
We emphasize, that it is crucial to consider the missing piece to obtain correct physical limits of the RG equation. When this is done for each contraction individually, factors will multiply to
\begin{equation*}
 (1+d\ell)^n=1+n\ d\ell +\mathcal{O}(d\ell^2)
\end{equation*}
with an general integer $n$. The \textit{total} missing piece will therefore (in general) not be equal to the term linear in $d\ell$, that contributes to the flow equation, but differ by a factor of $n$, which is an integer
corresponding to the number of performed contractions.

\section{Second order expansion}
\label{sec:AppB}
In second order, the focus is on the possibility of TPB. Correspondingly, time-ordering forces a rearrangement of the operators before normal-ordering,
\begin{align}\label{eq:FullProduct2}
& \mathcal{T} \pp \partial_x \theta_a(1)   e^{i\lambda \phi_a(1)}  \pp \times \pp \partial_x \theta_b(2)   e^{-i\lambda \phi_b(2)}  \pp\times \pp \partial_x \theta_c(3)  e^{i\lambda \phi_c(3)}  \pp\times \pp \partial_x \theta_d(4)   e^{-i\lambda \phi_d(4)}  \pp=\notag \\
& \pp \partial_x \theta_a(1)    e^{i\lambda \phi_a(1)}  \pp\times \pp \partial_x \theta_c(3)   e^{i\lambda \phi_c(3)}  \pp \times \pp \partial_x \theta_b(2)   e^{-i\lambda \phi_b(2)} \pp\times \pp \partial_x \theta_d(4)   e^{-i\lambda \phi_d(4)} \pp = \notag \\
& (2a (1+d\ell))^2 h(x-x')^2 \frac{1}{2^4} \pp  e^{i\lambda \phi_a(y_1, x)}  e^{i\lambda \phi_a(y_1,x')}\pp \pp  e^{-i\lambda \phi_b(y_2, x)}  e^{-i\lambda \phi_b(y_2, x')} \pp \bigg[s_0(x-x')^2 + \notag \\
&2 \lambda^2 s_0(x-x')(u(1,2)+\uT(1,2))^2+ (u_2(1,2)^2+\udT(1,2)^2)+2 \frac{\lambda^2}{4} (u_2(1,2)+\udT(1,2)) (u(1,2)+\uT(1,2))^2+ \notag \\
& \left(\frac{\lambda}{2} (u(1,2)+\uT(1,2))\right)^4 \bigg].
\end{align}
Both position variables $x, x'$ were kept while contracting $y_3\to y_1$ and $y_4\to y_2$.
Here, we introduced additional functions
\begin{align}
& h(x-x')=\left(\frac{2\pi}{L}\left|(x-x')+a\right| \right)^{\frac{\lambda^2}{2}}, \notag \\
& s_0(x-x')= \frac{2\left((1-\frac{\lambda^2}{2})a^2-(x-x')^2\right)}{(a^2+(x-x')^2)^2}, \notag \\
& \uT(1,2)=\frac{2(y+a)}{(y+a)^2+(x-x')^2}, \notag \\
&  \udT(1,2)=\frac{2\left((y+a)^2-(x-x')^2\right)}{\left((y+a)^2+(x-x')^2\right)^2}.\notag
\end{align}
In Eq.~(\ref{eq:FullProduct2}), the first term is expected to be the most important one, since all other terms decay with increasing time distances $y$. In a lowest order approximation, we therefore take into account only the first term,
identifying a TPB-contribution of the form
\begin{align}\label{eq:X22pFinalApp}
&  \exp \bigg[\int d\tau_1 d\tau_2\ \hat{\mathcal{H}}_{dis}(\tau_1, \tau_2)\bigg] \sim
  \frac{1}{2 a^4} \left(\frac{\DeT}{\pi^2 K} \right)^2 \left(\frac{2\pi a}{L}\right)^{8K} \sum_{a,b} \int dx dx' dy_1 dy_2 \ (1+2d\ell)\ m\left(\frac{x-x'}{a}\right) \notag \\
& \pp  e^{i\lambda \phi_a(y_1, x)}  e^{i\lambda \phi_a(y_1,x')}\pp \pp  e^{-i\lambda \phi_b(y_2, x)}  e^{-i\lambda \phi_b(y_2, x')} \pp +H.c.
\end{align}
The dependence on the spatial distance is embodied by a form factor
\begin{equation}
 m\left(\frac{x-x'}{a}\right)=\left(\frac{(1-2K)-\left(\frac{x-x'}{a}\right)^2}{\left(1+\left(\frac{x-x'}{a}\right)^2\right)^{2-K}}\right)^2.
\end{equation}
\clearpage

\twocolumngrid

\bibliography{LuttBib3}

\end{document}